\documentclass[conference, 10pt]{IEEEtran} 
\usepackage{graphicx}
\usepackage{epsfig}
\usepackage{epstopdf}
\usepackage{amsmath,amsthm,amsfonts,amssymb}
\usepackage{multirow}
\usepackage{cite}
\usepackage{tikz}
\usepackage{ellipsis}
\usepackage{xcolor}
\usepackage{xifthen}
\usepackage{cancel}
\usepackage{amsmath,bm}
\usepackage{array}
\usepackage[linesnumbered, ruled, vlined, noend]{algorithm2e}
\theoremstyle{definition} 
\theoremstyle{definition} 
\theoremstyle{definition} 
\theoremstyle{definition}

\newcommand{\B}{\beta}
\newcommand{\Lm}{\lambda}

\newcommand{\fixme}[2]{\ifx&#2&{\leavevmode\color{red}#1}\else{\leavevmode\color{red}FIXME\{}#1{\leavevmode\color{red}\}}\footnote{{\leavevmode\color{red}#2}}\PackageWarning{Fixme}{#1: #2}\fi}
\newcommand{\newstuff}[2]{\ifx&#2&{\leavevmode\color{blue}#1}\else{\leavevmode\color{blue}NEWSTUFF\{}#1{\leavevmode\color{blue}\}}\footnote{{\leavevmode\color{blue}#2}}\PackageWarning{Newstuff}{#1: #2}\fi}
\newcolumntype{M}[1]{>{\centering\arraybackslash}m{#1}}
\begin{document}

\title{Fast-SCAN decoding of Polar Codes}

\author{\IEEEauthorblockN{Charles Pillet, Carlo Condo, Valerio Bioglio}
\IEEEauthorblockA{Mathematical and Algorithmic Sciences Lab\\ Paris Research Center, Huawei Technologies Co. Ltd.\\
Email: $\{$charles.pillet1,carlo.condo,valerio.bioglio$\}$@huawei.com}} 

\maketitle

\begin{abstract}
Polar codes are able to achieve the capacity of memoryless channels under successive cancellation (SC) decoding. 
Soft Cancellation (SCAN) is a soft-output decoder based on the SC schedule, useful in iterative decoding and concatenation of polar codes. 
However, the sequential nature of this decoder leads to high decoding latency compared to state-of-the-art codes. 
To reduce the latency of SCAN, in this paper we identify special nodes in the decoding tree, corresponding to specific frozen-bit sequences, and propose dedicated low-latency decoding approaches for each of them.
The resulting fast-SCAN decoder does not alter the soft-output compared to the standard SCAN while dramatically reducing the decoding latency and yielding the same error-correction performance. 
\end{abstract}


\section{Introduction}
\label{sec:intro}
Polar codes are a class of linear block codes introduced in \cite{ArikanFirst}. 
They rely on the polarization effect to identify reliable and unreliable bit-channels, freezing the unreliable ones and transmitting information through the reliable ones. 
Polar codes can achieve capacity under successive cancellation (SC) decoding for infinite code length. 
However, for finite block length, SC yields poor error-correction performance and long decoding latency, due to its serial nature. 
On one side, to improve the performance of SC at minimal latency cost, SC list (SCL), and its evolution aided by cyclic-redundancy-check (CRC) have been proposed in \cite{TalSCL} and \cite{CRCaidedSCL}, respectively.
Concurrently, to reduce the latency of SC without any performance degradation, constituent codes that can be easily decoded have been investigated \cite{SSC,FastSC,FFastSC}. 
In particular, \cite{SSC} introduces a simplified-SC (SSC) decoder that can efficiently decode rate-0 and rate-1 nodes. 
Fast-SSC \cite{FastSC} improves SSC by implementing fast parallel decoders for single-parity-check nodes, repetition nodes and their mergers. 
More recently, \cite{FFastSC} identified 5 new types of nodes, providing their efficient decoders.

SCAN decoding \cite{SCANfirst} is a low-complexity soft output decoding algorithm based on the SC schedule, that can be effectively used in concatenation schemes and iterative decoding. 
Consequently, SCAN decoders suffer also from a poor latency due to the serial nature of the SC schedule. 
Moreover, SCAN allows iterative decoding, increasing its latency with the number of iterations. 
In order to speed up the decoding process, \cite{RCSCAN} introduces simplifications for rate-0 nodes (all frozen bits) and rate-1 nodes (all information bits). 

In this work, we detail fast decoders for several constituent codes under SCAN decoding improving the latency of SCAN without modifying its soft output, thus yielding the same extrinsic values for iterative message passing. 
An analysis of the decoding latency of the fast-SCAN decoder reveals up to $94.5\%$ reduction with respect to SCAN, while simulation results show that both decoders have the same error-correction performance.

\section{Preliminaries}
\label{sec:pre}
In this section we introduce the basic concepts on polar codes, together with the SC and SCAN decoding algorithms.

\subsection{Polar codes}\label{subsec:pc}
A polar code $(N,K)$ of length $N=2^n$ and dimension $K$ is a block code relying on the polarization effect of kernel matrix $G_2\triangleq \begin{bmatrix}
1 & 0\\
1& 1
\end{bmatrix}$. 
The polarization effect defined by transformation matrix $G_N = G_2^{\otimes n}$ creates $N$ virtual bit-channels, each one having a different reliability. 
The reliability of each channel may be computed through Density Evolution using Gaussian Approximation, the Bhattacharrya parameter or through Monte Carlo simulation \cite{frozenset}. 
The $K$ information bits are assigned to the $K$ most reliable bit-channels, while the remaining $N-K$ bit-channels are set to a known value, usually $0$, and represent the frozen set $\mathcal{F}$ of the code. 
The $N$-bit codeword $\bm{x}$ is generated as $\bm{x}=\bm{u}G_N$, where $\bm{u}$ is the input vector of the code, with $u_{i\in\mathcal{F}}=0$. 

\subsection{SC-based decoding}\label{subsec:SCdec}
\begin{figure}[t!]
  \centering
  \begin{tikzpicture}[scale=1.9, thick]
\newcommand\Triangle[1]{-- ++(0:2*#1) -- ++(120:2*#1) --cycle}
\newcommand\Square[1]{+(-#1,-#1) rectangle +(#1,#1)}

  \fill [gray, very thick] (0,0) circle [radius=.05];
  

  \fill [gray, very thick] (-1.0,-.5) circle [radius=.05]; 
  \fill [gray, very thick] (1,-.5) circle [radius=.05]; 

  \draw (-1.5,-1) circle [radius=.05];
  \fill [gray, very thick] (-.5,-1.0) circle [radius=.05];
 \fill [gray, very thick] (.5,-1) circle [radius=.05];
  \fill (1.5,-1) circle [radius=.05];

  \draw (-1.75,-1.5) circle [radius=.05];
  \draw (-1.25,-1.5) circle [radius=.05];
  \draw (-.75,-1.5) circle [radius=.05];
  \fill (-.25,-1.5) circle [radius=.05];
  \draw (.25,-1.5) circle [radius=.05];
  \fill (.75,-1.5) circle [radius=.05];
  \fill (1.25,-1.5) circle [radius=.05];
  \fill (1.75,-1.5) circle [radius=.05];

  \node at (-1.75,-1.7) {$\hat{u}_0$};
  \node at (-1.25,-1.7) {$\hat{u}_1$};
  \node at (-.75,-1.7) {$\hat{u}_2$};
  \node at (-.25,-1.7) {$\hat{u}_3$};
  \node at (.25,-1.7) {$\hat{u}_4$};
  \node at (.75,-1.7) {$\hat{u}_5$};
  \node at (1.25,-1.7) {$\hat{u}_6$};
  \node at (1.75,-1.7) {$\hat{u}_7$};

  \draw (0,-.05) -- (-1,-.45);
  \draw (0,-.05) -- (1,-.45);

  \draw (-1,-.55) -- (-1.5,-.95);
  \draw (-1,-.55) -- (-.5,-.95);
  \draw (1,-.55) -- (.5,-.95);
  \draw (1,-.55) -- (1.5,-.95);

  \draw (-1.5,-1.05) -- (-1.75,-1.45);
  \draw (-1.5,-1.05) -- (-1.25,-1.45);
  \draw (-.5,-1.05) -- (-.75,-1.45);
  \draw (-.5,-1.05) -- (-.25,-1.45);
  \draw (.5,-1.05) -- (.25,-1.45);
  \draw (.5,-1.05) -- (.75,-1.45);
  \draw (1.5,-1.05) -- (1.25,-1.45);
  \draw (1.5,-1.05) -- (1.75,-1.45);

  \draw [very thin,gray,dashed] (-2,0) -- (2,0);
  \draw [very thin,gray,dashed] (-2,-.5) -- (2,-.5);
  \draw [very thin,gray,dashed] (-2,-1) -- (2,-1);
  \draw [very thin,gray,dashed] (-2,-1.5) -- (2,-1.5);

  \node at (-2.25,0) {$t=3$};
  \node at (-2.25,-.5) {$t=2$};
  \node at (-2.25,-1) {$t=1$};
  \node at (-2.25,-1.5) {$t=0$};

  \draw [->] (-.12,-.05) -- (-1,-.4) node [above=-0.1cm,midway,rotate=25] {$\bm{\lambda}$};
  \draw [->] (-.88,-.45) -- (0,-.1) node [below=-0.1cm,midway,rotate=25] {$\bm{\beta}$};
  \draw [->] (-1.06,-.55) -- (-1.5,-.9) node [above=.1mm,midway] {$\bm{\lambda}^{\ell}$};
  \draw [->] (-1.44,-.95) -- (-1.0,-0.6) node [below=-.1cm,midway] {$\bm{\beta}^{\ell}$};

  \draw [<-] (-.94,-.55) -- (-.5,-.9);
  \draw [<-] (-.56,-.95) -- (-0.975,-.625);
  \node at (-0.85, -0.875) {$\bm{\lambda}^{\text{r}}$};
  \node at (-0.66, -0.6) {$\bm{\beta}^{\text{r}}$};

\end{tikzpicture}
  \caption{SC decoder of an $(8,4)$ polar code, $\mathcal{F} = \{0,1,2,4\}$.}
  \label{fig:treeSC}
\end{figure}
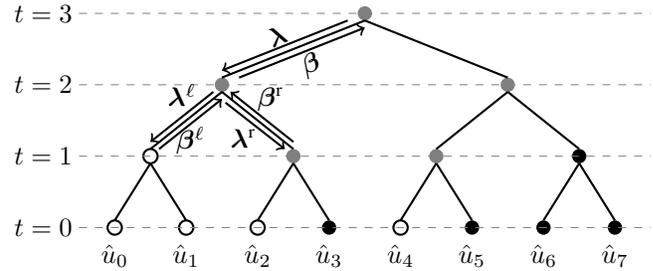
SC has been proposed in \cite{ArikanFirst} as the native polar code decoding algorithm. 
As shown in Figure~\ref{fig:treeSC}, it can be described as a binary tree search, where the tree is traversed depth-first starting from the left branch.  
We consider the soft information received as input from the channel to be in the form of logarithmic likelihood ratios (LLRs). 
A node stage $t$ represents a constituent code of length $2^t$, receiving from its parent node the soft information vector $\bm{\Lm}$ of length $2^t$. 
This is used to compute the $2^{t-1}$-element soft information vector $\bm{\Lm}_\ell$ for its left child. 
The left child eventually returns its constituent codeword $\bm{\B}_\ell$, that is used to compute the soft information $\bm{\Lm}_r$ to be sent to the right child. 
After receiving the second constituent codeword $\bm{\B}_r$ from the right child, the node feeds back the $2^t$-element estimated codeword $\bm{\B}$ to its parent node. 

In order to reduce the latency of SC-based decoders, the SSC decoder \cite{SSC} is able to decode sub-trees constituted of information bits (rate-1) or frozen bits (rate-0) without having to explore them. 
Later, single-parity-check (SPC) and repetition (REP) nodes have been decoded efficiently in \cite{FastSC} to further speed up SC decoding. 
Finally, five new nodes were identified in \cite{FFastSC}, while generalized and expanded versions have been proposed in \cite{gen_fastSC}. 

\subsection{SCAN decoding}\label{subsec:scan}
\begin{figure}[t!]
\centering
\begin{tikzpicture}[scale=1.9, thick]
\newcommand\Triangle[1]{-- ++(0:2*#1) -- ++(120:2*#1) --cycle}
\newcommand\Square[1]{+(-#1,-#1) rectangle +(#1,#1)}

  \fill [gray, very thick] (-1,0) circle [radius=.05];
  
  
  \draw [very thin,gray,dashed] (-2,0) -- (0.1,0);
  \draw [very thin,gray,dashed] (-2,-.5) -- (0.1,-0.5);
  \draw [very thin,gray,dashed] (-2,-1) -- (0.1,-1);

  \fill [red, very thick] (-1.0,-.5) circle [radius=.05]; 

  \fill [gray, very thick] (-1.75,-1) circle [radius=.05];
  \fill [gray, very thick] (-0.25,-1) circle [radius=.05];

  \draw (-1,-0.05) -- (-1,-.45);
  
  \draw (-1,-.55) -- (-1.75,-0.95);
  \draw (-1,-.55) -- (-0.25,-0.95);

  \node at (-2.25,0) {$t+1$};
  \node at (-2.25,-.5) {$t$};
  \node at (-2.25,-1) {$t-1$};

  \draw [->] (-1.05,-0.05) -- (-1.05,-.45) node [left=-.08cm,midway] {$\bm{\lambda}^i_t$};
  \draw [->] (-.95,-.45) -- (-0.95,-0.05) node [right=-.08cm,midway] {$\bm{\beta}^i_t$};

  \draw [->] (-1.08,-.55) -- (-1.75,-0.9) node [above=-.1cm,midway,rotate=30] {$\bm{\lambda}^{2i}_{t-1}$};
  \draw [->] (-1.69,-0.97) -- (-1.0,-0.6) node [below=-.1cm,midway,rotate=30] {$\bm{\beta}^{2i}_{t-1}$};

  \draw [<-] (-.92,-.55) -- (-0.25,-0.9) node [above=-.1cm,midway,rotate=-30] {$\bm{\beta}^{2i+1}_{t-1}$};
  \draw [<-] (-0.31,-0.97) -- (-0.975,-.625) node [below=-.1cm,midway,rotate=-30] {$\bm{\lambda}^{2i+1}_{t-1}$};

\end{tikzpicture}
\caption{Soft message exchange between decoding tree stages.}
\label{fig:scan2}
\end{figure}
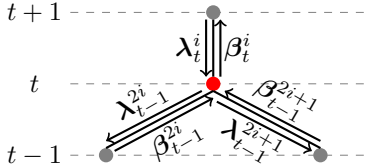
SCAN decoding \cite{SCANfirst} relies on the SC schedule to exchange soft information through the decoding tree in both directions. 
This allows to refine the soft information both at the root and at the leaves of the tree by iterating the decoding process. 
Compared to SC, SCAN returns soft values instead of hard decisions, slightly increasing decoding complexity and latency.

Stage $t$ of the decoding tree is constituted of $2^{n-t}$ nodes of size $2^t$, where $0 \leq t \leq n$. 
The $i$-th node at stage $t$ receives vector $\bm{\Lm}_{t}^{i}$ including $2^t$ LLRs from its parent node, and performs update operations to feed back the $2^t$-element soft vector $\bm{\B}_{t}^{i}$. 
As with SC, the vector $\Lm_{n}^{0}$ is initialized with channel LLRs. 
Moreover, the decoder can exploit \emph{a priori} information coming from the frozen set $\mathcal{F}$;  
the message $\bm{\B}_{0}^{i}$ fed back from the leaves, corresponding to the estimated vector $\mathbf{\hat{u}}$, is set to 
\begin{equation}\label{eq:Binit}
\B_0^{i} =
  \begin{cases}
    \infty  & \text{if } i \in \mathcal{F}\text{,}\\
    0  & \text{otherwise.}
  \end{cases}
\end{equation}
The other messages are initialized to $0$ since no further \emph{a priori} information is available. 
It is worth noting that message sets $\bm{\B}_{0}^{i}$ and $\Lm_{n}^{0}$ are not updated throughout the decoding and keep their initial values. 

Message passing through the decoding tree follows the SC scheduling described in Section \ref{subsec:SCdec}. 
Figure~\ref{fig:scan2} represents a node at stage $t$, i.e. a constituent code of length $2^t$. 
First, the soft message vector to be sent to the left child is computed as  
\begin{align}\label{eq:lambdaI}
\Lm^{2i}_{t-1}[k]=\tilde{f}\left(\Lm_{t}^{i}[k],\Lm_{t}^{i}[k+2^{t-1}]+\B_{t-1}^{2i+1}[k]\right)
\end{align}
for $k =\left\{0,\dots,2^{t-1}-1\right\}$. 
Then, the node receives the soft message vector $\bm\B^{2i}_{t-1}$ from its left child and computes the message $\bm\Lm^{2i+1}_{t-1}$ sent to its right child as
\begin{align}\label{eq:lambdaII}
\Lm^{2i+1}_{t-1}[k]=\tilde{f}\left(\Lm_{t}^{i}[k],\B_{t-1}^{2i}[k]\right)+\Lm_{t}^{i}[k+2^{t-1}]. 
\end{align}
As soon as the soft message vector $\bm{\B}^{2i+1}_{t-1}$ is received from the right child node, the feedback soft message vector $\bm{\B}_t^{i}$ of length $2^t$ is calculated as
\begin{align}
\B_t^{i}[k]&=\tilde{f}(\B_{t-1}^{2i}[k],\Lm_{t}^{i}[k+2^{t-1}]+\B_{t-1}^{2i+1}[k]) \label{eq:betaI}\\
\B_t^{i}[k+2^{t-1}]&=\B_{t-1}^{2i+1}[k]+\tilde{f}(\Lm_{t}^{i}[k],\B_{t-1}^{2i}[k]) \label{eq:betaII}
\end{align}
and sent to the parent node. 
Function $\tilde{f}: \mathbb{R}^2 \rightarrow \mathbb{R}$ is the box-plus operator
\begin{equation}\label{eq:fftilde}
	\tilde{f}(a,b) = a \boxplus b \triangleq \log\left(\frac{1+e^{a+b}}{e^a+e^b}\right)
\end{equation}
whose hardware-friendly implementation is given by 
\begin{equation}
	\tilde{f}(a,b) \simeq \text{min}\left(\left|a\right|,\left|b\right|\right)\text{sign}(a)\text{sign}(b). 
\end{equation}

\section{Fast-SCAN decoding}
\label{sec:fast-scan}
\begin{figure}[t!]
  \centering
  \resizebox{0.485\textwidth}{!}{\begin{tikzpicture}[scale=0.9, thick]
\newcommand\Triangle[1]{-- ++(120:2*#1) -- ++(0:2*#1) --cycle}
\newcommand\Triangledroit[1]{-- ++(0:2*#1) -- ++(120:2*#1) --cycle}
\newcommand\Square[1]{+(-#1,-#1) rectangle +(#1,#1)}
\def\scale{2}

  \draw [very thin,gray,dashed] (-2*\scale,0) -- (2*\scale,0);
  \draw [very thin,gray,dashed] (-2*\scale,-.5) -- (2*\scale,-.5);
  \draw [very thin,gray,dashed] (-2*\scale,-1) -- (2*\scale,-1);
  \draw [very thin,gray,dashed] (-2*\scale,-1.5) -- (2*\scale,-1.5);
  \draw [very thin,gray,dashed] (-2*\scale,-2) -- (2*\scale,-2);
  \draw [very thin,gray,dashed] (-2*\scale,-2.5) -- (2*\scale,-2.5);
  \draw [very thin,gray,dashed] (-2*\scale,-3) -- (2*\scale,-3);

  \node at (-2.5*\scale,0) {$t=8$};
  \node at (-2.5*\scale,-.5) {$t=7$};
  \node at (-2.5*\scale,-1) {$t=6$};
  \node at (-2.5*\scale,-1.5) {$t=5$};
  \node at (-2.5*\scale,-2) {$t=4$};
  \node at (-2.5*\scale,-2.5) {$t=3$};
  \node at (-2.5*\scale,-3) {$t=2$};

  \draw [black, very thin ,fill=gray] (0,0) circle [radius=.05];
  

  \draw [black, very thin ,fill=gray] (-1.0*\scale,-.5) circle [radius=.05]; 
  \draw [black, very thin ,fill=red] (1*\scale,-.5) \Square{.05};

  \draw [black, very thin ,fill=gray] (-1.5*\scale,-1) circle [radius=.05];
  \draw [black, very thin ,fill=red] (-.5*\scale,-1.0) \Square{.05};
  
  \draw [black, very thin ,fill=gray] (-1.75*\scale,-1.5) circle [radius=.05];
  \draw [black, very thin ,fill=red] (-1.25*\scale,-1.5) \Square{.05};

  \draw [black, very thin ,fill=gray] (-1.875*\scale,-2) circle [radius=.05];
  \draw [black, very thin ,fill=gray] (-1.625*\scale,-2) circle [radius=.05];

  \draw [black, very thin ,fill=blue!40] (-1.9375*\scale,-2.5) \Square{.05};
  \draw [black, very thin ,fill=gray] (-1.8125*\scale,-2.5) circle [radius=.05];
  \draw [black, very thin ,fill=gray] (-1.6875*\scale,-2.5) circle [radius=.05];
  \draw [black, very thin ,fill=black] (-1.55*\scale,-2.5) circle [radius=.05];

  \draw [black, very thin ,fill=blue!40] (-1.84375*\scale,-3) \Square{.05};
  \draw [black, very thin, fill=red] (-1.78125*\scale,-3) \Square{.05};
  \draw [black, very thin ,fill=blue!40] (-1.7175*\scale,-3.) \Square{.05};
  \fill [black, very thick] (-1.65625*\scale,-3) circle [radius=.05];

  \draw [thin] (0,-.05) -- (-1*\scale,-.45);
  \draw [thin] (0,-.05) -- (1*\scale,-.45);

  \draw [thin] (-1*\scale,-.55) -- (-1.5*\scale,-.95);
  \draw [thin] (-1*\scale,-.55) -- (-.5*\scale,-.95);

  \draw [thin] (-1.5*\scale,-1.05) -- (-1.75*\scale,-1.45);
  \draw [thin] (-1.5*\scale,-1.05) -- (-1.25*\scale,-1.45);
  
 \draw [thin] (-1.75*\scale,-1.55) -- (-1.875*\scale,-1.95);
 \draw [thin] (-1.75*\scale,-1.55) -- (-1.625*\scale,-1.95);
 
 \draw [thin] (-1.875*\scale,-2.05) -- (-1.8125*\scale,-2.45);
 \draw [thin] (-1.875*\scale,-2.05) -- (-1.9375*\scale,-2.45);
 \draw [thin] (-1.625*\scale,-2.05) -- (-1.6875*\scale,-2.45);
 \draw [thin] (-1.625*\scale,-2.05) -- (-1.55*\scale,-2.45);

 \draw [thin] (-1.8125*\scale,-2.55) -- (-1.84375*\scale,-2.95);
 \draw [thin] (-1.8125*\scale,-2.55) -- (-1.78125*\scale,-2.95);
 \draw [thin] (-1.6875*\scale,-2.55) -- (-1.7175*\scale,-2.95);
 \draw [thin] (-1.6875*\scale,-2.55) -- (-1.65625*\scale,-2.95);

\end{tikzpicture}}
	  \caption{Decoding tree of fast-SCAN for (256,239) component codes;  red squares are SPC nodes, black circles are rate-1 nodes and light blue squares are repetition nodes.}
  \label{fig:Fast256239}
\end{figure}
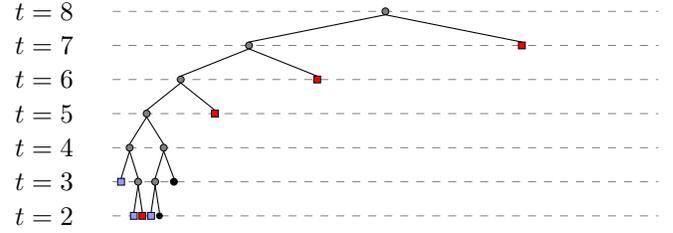
In this section we introduce SCAN message update rules for several special nodes used in fast SC decoders. 
Fast-SCAN can provide the same soft output of SCAN while considerably reducing the decoding latency.
Figure~\ref{fig:Fast256239} depicts the pruned decoding tree as explored by fast-SCAN decoding for the $(256,239)$ polar code designed according to the 5G standard \cite{5G}. 
The full tree would be composed by 511 nodes, that are reduced to 17 through constituent code fast decoding. 
%

Simplified SCAN (SSCAN) decoding has been proposed in \cite{RCSCAN}, where decoders for rate-0 and rate-1 nodes are examined. 
A rate-0 node always returns $\bm{\B}^{i}_{t}=\left[\infty,\infty,\dots,\infty\right]$ regardless of the input LLRs. 
Instead, for rate-1 nodes, the SCAN decoder returns the all-zero vector $\bm{\B}^{i}_{t}=\left[0,0,\dots,0\right]$. 

\subsection{SPC nodes}\label{subsec:SPC}
In an SPC node the first leaf is frozen while all the other leaves represent information bits. 
SPC nodes are more likely to occur in high-rate polar codes \cite{FastSC}, that are used in constructions such as product polar codes \cite{PPC}. 
The SPC node imposes an even parity constraint on its bits; the Fast-SSC \cite{FastSC} algorithm decodes SPC nodes with Wagner decoding \cite{Wagner}, i.e. by flipping the least reliable bit if the overall parity is not satisfied. 

When we decode an SPC node with SCAN, the $\beta$ term in \eqref{eq:lambdaI} is equal to 0 until $t\geq2$, hence, entry $k$ of $\Lm^{2i}_{t-1}$ is the box-plus operation of $\Lm^{i}_{t}[k]$ and $\Lm^{i}_{t}[k+2^{t-1}]$. By induction at stage 1, $\Lm^{2^{t-1}i}_{1}[0]$ and $\Lm^{2^{t-1}i+1}_{1}[1]$ are respectively the box-plus operation of all even-indexed and odd-indexed $\Lm^{i}_{t}$ values. The frozen set imposes $\B_{0}^{2^{t}i}=\{\infty\}$ and $\B_{0}^{2^{t}i+1}=\{0\}$ and by using \eqref{eq:betaI}-\eqref{eq:betaII}, $\B_{1}^{2^{t-1}i}=\left\{\Lm_{1}^{2^{t-1}i}[1],\Lm_{1}^{2^{t-1}i}[0]\right\}$. 
The structure of SPC nodes guarantees that the feedback from the right sub-tree is always the all-zero vector; given \eqref{eq:betaI}-\eqref{eq:betaII},{} entry $k$ of $\bm{\B}^{i}_{t}$ is the result of the box-plus operation of all entries of $\bm{\Lm}^{i}_{t}$ excluding $\Lm^{i}_{t}[k]$:
\begin{align}\label{eq:resultSPCSCAN}
	\B^{i}_{t}[k] &= \overset{2^t-1}{\underset{j=0, j\neq k}{\boxplus}}\Lm^{i}_{t}[j] = \\
 	\nonumber &\simeq \underset{0 \leq j < 2^t, j\neq k}{\text{min}}\left(\left|\Lm^{i}_{t}[j]\right|\right)\prod^{2^t-1}_{j=0, j\neq k} \text{sign}(\Lm^{i}_{t}[j])~. 
\end{align}
We can thus decode SPC nodes without traversing the tree. 
We can rewrite \eqref{eq:resultSPCSCAN} as 
\begin{equation}\label{eq:SPCSCAN}
\B^{i}_{t}[k]=\left\{
  \begin{array}{@{}ll@{}}
    \left(-1\right)^{P\oplus h[k_0]}\left|\Lm^{i}_{t}[k_1]\right| & \text{if } k = k_0~,\\
    \left(-1\right)^{P\oplus h[k]}   \left|\Lm^{i}_{t}[k_0]\right| & \text{otherwise~,}
  \end{array}\right.
\end{equation} 
where $P$ is the overall parity 
\begin{equation}\label{eq:xorHD}
	P = \overset{2^t-1}{\underset{j=0}{\bigoplus}} h[j]~,
\end{equation}
$h$ is the hard decision taken on the LLRs in $\bm{\Lm}^{i}_{t}$, and $k_{1,2}$ are the indices of the the least and second least reliable values of $\bm{\Lm}^{i}_{t}$.
It is worth noticing that \eqref{eq:SPCSCAN} does not change the parity of the input LLRs, while on the contrary Wagner decoding forces the even parity constraint. 
When an SPC node feeds back a vector having a wrong parity, SCAN decoding may fail due to the extrinsic nature of its output.
In order to force the parity condition while keeping the expected LLRs distribution \eqref{eq:SPCSCAN} can be modified as
\begin{equation}\label{eq:proposeSPCSCAN}
  \B^{i}_{t}[k]=\left\{
  \begin{array}{@{}ll@{}}
    \left(-1\right)^{P\oplus h[k_0]}\left|\Lm^{i}_{t}[k_1]\right| & \text{if } k = k_0~,\\
    \text{sign}\left(\Lm^{i}_{t}[k]\right) \cdot \left|\Lm^{i}_{t}[k_0]\right| & \text{otherwise~.}
  \end{array}\right.
\end{equation} 
However, the output will be no longer extrinsic, and will differ from that of SCAN. 
In the following, we propose to use \eqref{eq:SPCSCAN} considering that a key application of fast-SCAN is the speed-up of iterative decoding of polar-based code constructions. 

\subsection{REP nodes}\label{subsec:REP}

A REP node occurs when all the leaves of a node are frozen except the rightmost one. 
As a result, the value of the $2^t$ bits at the root of the REP node is equal to the information bit. 
Fast-SSC decodes REP nodes through hard decision on the sum of all the elements in $\bm{\Lm}^{i}_{t}$. 
The result is then replicated $2^t$ times in the feedback.
Concerning the SCAN decoder, the $\beta$ term in \eqref{eq:lambdaII} representing the feedback from the left sub-tree is always a vector of infinitives for $t\geq2$, leading to $\Lm^{2i+1}_{t-1}[k]=\Lm^{i}_{t}[k]+\Lm^{i}_{t}[k+2^{t-1}]$. 
At stage 1, $\Lm^{2^{t-1}(i+1)-1}_{1}[0]$ and $\Lm^{2^{t-1}(i+1)-1}_{1}[1]$ are respectively the sum of all even-indexed and odd-indexed $\Lm^{i}_{t}$ values. 
Similarly to the SPC case, $\B_{0}^{2^{t}(i+1)-2}=\{\infty\}$ and $\B_{0}^{2^{t}(i+1)-1}=\{0\}$, and thus $\B_{1}^{2^{t-1}(i+1)-1}=\left\{\Lm_{1}^{2^{t-1}(i+1)-1}[1],\Lm_{1}^{2^{t-1}(i+1)-1}[0]\right\}$. 
Finally, each entry $k$ of $\bm{\B}^{i}_{t}$ can be computed without traversing the tree as the sum all the entries of $\bm{\Lm}^{i}_{t}$ excluding $k$: 
\begin{equation}\label{eq:rep}
	\B^{i}_{t}[k] = \sum_{j=0}^{2^t-1}\Lm^{i}_{t}[j] - \Lm^{i}_{t}[k].  
\end{equation}

\subsection{Type-X nodes}\label{subsec:type}
In \cite{FFastSC}, the authors provide 5 new nodes, named Type-I to Type-V.
A type-I node, also known as REP-II node, has all leaves frozen except the last two. 
They can be decoded as separate REP nodes identified by even-indexed and odd-indexed bits. 
Consequently fast-SCAN computes element $k$ of $\B^{i}_{t}$, with $k = 0,\dots,2^{t-1}-1$, as 
\begin{align}
	\B^{i}_{t}[2k] &= \sum_{j=0, j\neq k}^{2^{t-1}-1}\Lm^{i}_{t}[2j]~, \\
	\B^{i}_{t}[2k+1] &= \sum_{j=0, j\neq k}^{2^{t-1}-1}\Lm^{i}_{t}[2j+1]~,
\end{align}
without traversing the type-I tree.

Type-III nodes have 2 frozen bits located on the first two bit-channels, while the other bit-channels are unfrozen . 
They may be decoded as two separate SPC nodes composed by the even-indexed and odd-indexed values. 
We denote $k_{0,e}$ and $k_{0,o}$ the least reliable indices corresponding to the set of even-indexed and odd-indexed bits, with $k_{1,e}$ and $k_{1,o}$ being the second least reliable index of each set. 
The overall parities $P_e$ and $P_o$ are computed in order to calculate the soft message feedback as in \eqref{eq:SPCSCAN}:
\begin{align*}
 \B^{i}_{t}[2k]&=\left\{
  \begin{array}{@{}ll@{}}
    \left(-1\right)^{P_e\oplus h[k_{0,e}]}\left|\Lm^{i}_{t}[k_{1,e}]\right| & \text{if } 2k = k_{0,e}\\
    \left(-1\right)^{P_e\oplus h[2k]}\left|\Lm^{i}_{t}[k_{0,e}]\right| & \text{otherwise.}
  \end{array}\right. \\
   \B^{i}_{t}[2k+1]&=\left\{
  \begin{array}{@{}ll@{}}
    \left(-1\right)^{P_o\oplus h[k_{0,o}]}\left|\Lm^{i}_{t}[k_{1,o}]\right| & \text{if } 2k+1 = k_{0,o}\\
    \left(-1\right)^{P_o\oplus h[2k+1]}\left   |\Lm^{i}_{t}[k_{0,o}]\right| & \text{otherwise.}
  \end{array}\right.
\end{align*} 

The other type-X nodes presented in \cite{FFastSC} can be decoded without traversing the tree as well. For Type-II and Type-IV nodes, entry $k$ of $\B_t^i$ is computed as:
{\scriptsize
\begin{align}
\B^{i}_{t}[k]&=\nonumber&\overset{3}{\underset{j=0, j\neq k[4]}{\boxplus}}\sum_{m=0}^{2^{t-2}-1}\Lm^{i}_{t}[4m+j]
+\sum_{m=0,4m+k[4]\neq k}^{2^{t-2}-1}\Lm^{i}_{t}[4m+k[4]]~, \\
\B^{i}_{t}[k]&=\nonumber&\sum_{j=0, j\neq k[4]}^{3}\overset{2^{t-2}-1}{\underset{m=0}{\boxplus}}\Lm^{i}_{t}[4m+j] \boxplus\overset{2^{t-2}-1}{\underset{m=0, 4m+k[4]\neq k}{\boxplus}}\Lm^{i}_{t}[4m+k[4]]~,
\end{align}
}%
i.e. a combination of sum and box-plus operations among values of $\Lm_t^i$ selected with modulo-4 indexing. While these equations do not alter the soft output of SCAN, their computational complexity is substantially higher than the other identified nodes, and they will not be considered in the following Sections. 
Finally, the structure of Type-V nodes implies a frozen bit embedded in a series of information bits. As a consequence, aside from its high complexity, the mathematical expression for the computation of $\B^{i}_{t}$ at the root is only valid for the first iteration. Thus, Type-V nodes are not considered in the remainder of the paper either.

\section{Decoding Latency Analysis}\label{sec:latency}
%
In this Section we evaluate the decoding latency of the proposed Fast-SCAN decoder and compare it to the latency of the standard SCAN decoder. 
Similarly to \cite{SSC,FFastSC}, we suppose that hard decisions on LLRs and bit operations are executed instantaneously, while operations involving real numbers (additions, comparisons) and Wagner decoding require one clock cycle.
With this assumption, one SCAN update rule consumes 2 clock cycles, since it is composed of a box-plus operation followed by an addition, as discussed in Section~\ref{subsec:scan}. 
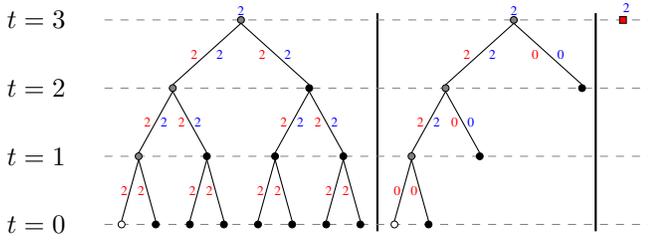
\begin{figure}[t!]
  \centering
  \resizebox{0.485\textwidth}{!}{\begin{tikzpicture}[scale=0.9, thick]
\newcommand\Triangle[1]{-- ++(120:2*#1) -- ++(0:2*#1) --cycle}
\newcommand\Triangledroit[1]{-- ++(0:2*#1) -- ++(120:2*#1) --cycle}
\newcommand\Square[1]{+(-#1,-#1) rectangle +(#1,#1)}
\def\scale{2}

  \draw [black] (0,0.1) -- (0,-3.1);  
  \draw [black] (1.6*\scale,0.1) -- (1.6*\scale,-3.1);  
  
  \draw [very thin,gray,dashed] (-2*\scale,0) -- (2*\scale,0);
  \draw [very thin,gray,dashed] (-2*\scale,-1) -- (2*\scale,-1);
  \draw [very thin,gray,dashed] (-2*\scale,-2) -- (2*\scale,-2);
  \draw [very thin,gray,dashed] (-2*\scale,-3) -- (2*\scale,-3);

  \node at (-2.5*\scale,0) {$t=3$};
  \node at (-2.5*\scale,-1) {$t=2$};
  \node at (-2.5*\scale,-2) {$t=1$};
  \node at (-2.5*\scale,-3) {$t=0$};

  \draw [black, thin ,fill=gray] (-1.0*\scale,0) circle [radius=.05] node [above=-0.6mm, text=blue] {\tiny{2}};
  \draw [black, thin ,fill=gray] (1.0*\scale,0) circle [radius=.05] node [above=-0.6mm, text=blue] {\tiny{2}};
  \draw [black, very thin ,fill=red] (1.8*\scale,0) \Square{0.05} node [above=-0.6mm, text=blue] {\tiny{2}};
    \draw [thin] (-1*\scale,-.05) -- (-0.5*\scale,-.95) node[midway, right, text=blue] {\tiny{2}} node[midway, left, text=red] {\tiny{2}};
  \draw [thin] (-1*\scale,-.05) -- (-1.5*\scale,-.95) node[midway, right, text=blue] {\tiny{2}} node[midway, left, text=red] {\tiny{2}};
  \draw [black, very thin ,fill=gray] (-1.5*\scale,-1) circle [radius=.05]; 
  \draw [black, very thin ,fill=black] (-0.5*\scale,-1) circle [radius=.05];
  
  \draw [thin] (-1.5*\scale,-1.05) -- (-1.75*\scale,-1.95) node[midway, right=-0.6mm, text=blue] {\tiny{2}} node[midway, left=-0.6mm, text=red] {\tiny{2}};
  \draw [thin] (-1.5*\scale,-1.05) -- (-1.25*\scale,-1.95) node[midway, right=-0.6mm, text=blue] {\tiny{2}} node[midway, left=-0.6mm, text=red] {\tiny{2}};
  \draw [thin] (-0.5*\scale,-1.05) -- (-0.75*\scale,-1.95) node[midway, right=-0.6mm, text=blue] {\tiny{2}} node[midway, left=-0.6mm, text=red] {\tiny{2}};
  \draw [thin] (-0.5*\scale,-1.05) -- (-0.25*\scale,-1.95) node[midway, right=-0.6mm, text=blue] {\tiny{2}} node[midway, left=-0.6mm, text=red] {\tiny{2}};
  \draw [black, very thin ,fill=gray] (-1.75*\scale,-2) circle [radius=.05];
  \draw [black, very thin ,fill=black] (-1.25*\scale,-2) circle [radius=.05];
  \draw [black, very thin ,fill=black] (-0.75*\scale,-2) circle [radius=.05];
  \draw [black, very thin ,fill=black] (-0.25*\scale,-2) circle [radius=.05];

  \draw [thin] (-1.75*\scale,-2.05) -- (-1.875*\scale,-2.95) node[midway, left=-0.9mm, text=red] {\tiny{2}};
  \draw [thin] (-1.75*\scale,-2.05) -- (-1.625*\scale,-2.95) node[midway, left=-0.9mm, text=red] {\tiny{2}};
  \draw [thin] (-1.25*\scale,-2.05) -- (-1.375*\scale,-2.95) node[midway, left=-0.9mm, text=red] {\tiny{2}};
  \draw [thin] (-1.25*\scale,-2.05) -- (-1.125*\scale,-2.95) node[midway, left=-0.9mm, text=red] {\tiny{2}};
  \draw [thin] (-0.75*\scale,-2.05) -- (-0.875*\scale,-2.95) node[midway, left=-0.9mm, text=red] {\tiny{2}};
  \draw [thin] (-0.75*\scale,-2.05) -- (-0.625*\scale,-2.95) node[midway, left=-0.9mm, text=red] {\tiny{2}};
  \draw [thin] (-0.25*\scale,-2.05) -- (-0.375*\scale,-2.95) node[midway, left=-0.9mm, text=red] {\tiny{2}};
  \draw [thin] (-0.25*\scale,-2.05) -- (-0.125*\scale,-2.95) node[midway, left=-0.9mm, text=red] {\tiny{2}};
  
  \draw [black, very thin ,fill=white] (-1.875*\scale,-3) circle [radius=.05];
  \draw [black, very thin ,fill=black] (-1.375*\scale,-3) circle [radius=.05];
  \draw [black, very thin ,fill=black] (-1.625*\scale,-3) circle [radius=.05];
  \draw [black, very thin ,fill=black] (-1.125*\scale,-3) circle [radius=.05];
  \draw [black, very thin ,fill=black] (-0.875*\scale,-3) circle [radius=.05];
  \draw [black, very thin ,fill=black] (-0.375*\scale,-3) circle [radius=.05];
  \draw [black, very thin ,fill=black] (-0.625*\scale,-3) circle [radius=.05];
  \draw [black, very thin ,fill=black] (-0.125*\scale,-3) circle [radius=.05];

  \draw [black, very thin ,fill=black] (1.5*\scale,-1) circle [radius=.05]; 
  \draw [black, very thin ,fill=gray] (0.5*\scale,-1) circle [radius=.05];

  \draw [thin] (1*\scale,-0.05) -- (0.5*\scale,-0.95) node[midway, right, text=blue] {\tiny{2}} node[midway, left, text=red] {\tiny{2}};
  \draw [thin] (1*\scale,-0.05) -- (1.5*\scale,-0.95) node[midway, right, text=blue] {\tiny{0}} node[midway, left, text=red] {\tiny{0}};
  \draw [black, very thin ,fill=gray] (0.25*\scale,-2) circle [radius=.05];
  \draw [black, very thin ,fill=black] (0.75*\scale,-2) circle [radius=.05];
  
  \draw [thin] (0.5*\scale,-1.05) -- (0.75*\scale,-1.95) node[midway, right=-0.6mm, text=blue] {\tiny{0}} node[midway, left=-0.6mm, text=red] {\tiny{0}};
  \draw [thin] (0.5*\scale,-1.05) -- (0.25*\scale,-1.95) node[midway, right=-0.6mm, text=blue] {\tiny{2}} node[midway, left=-0.6mm, text=red] {\tiny{2}};
  
  \draw [black, very thin ,fill=white] (0.125*\scale,-3) circle [radius=.05];
  \draw [black, very thin ,fill=black] (0.375*\scale,-3) circle [radius=.05];
  
  \draw [thin] (0.25*\scale,-2.05) -- (0.375*\scale,-2.95) node[midway, left=-0.9mm, text=red] {\tiny{0}};
  \draw [thin] (0.25*\scale,-2.05) -- (0.125*\scale,-2.95) node[midway, left=-0.9mm, text=red] {\tiny{0}};

\end{tikzpicture}}
  \caption{SCAN, Simplified SCAN and Fast-SCAN decoding tree of the SPC node, with relative latencies.}
  \label{fig:dectreeSCAN}
\end{figure}

Figure~\ref{fig:dectreeSCAN} depicts the decoding trees of the three decoders for a constituent code of length $N_v=8$. $N_v-2$ edges have a latency of 4 clock cycles, i.e. 2 for the computation of vector $\bm{\Lm}$ \eqref{eq:lambdaI}-\eqref{eq:lambdaII} and 2 for vector $\bm{\B}$ \eqref{eq:betaI}-\eqref{eq:betaII}. At the same time, $N_v$ edges at stage $0$ have a latency of 2 clock cycles, since $\bm{\B}_{0}^{i}$ are not updated through the decoding. 
Finally, the root has a latency of 2 corresponding to the computation of the vector $\bm{\B}_{n}^{i}$.
The overall SCAN latency for a node at stage $t$ is then $L=4*(N_v-2)+2*N_v+2=6\cdot(2^{t}-1)$.



As seen in Figure~\ref{fig:dectreeSCAN}, the latency of SSCAN to decode an SPC node of size $2^t$ is $L=4*(t-1)+2$, since the constant values of $\bm{\B}_{t}^{i}$ allow for instantaneous decoding of rate-0 and rate-1 nodes.
For fast-SCAN, hard decisions and \eqref{eq:xorHD} are executed instantaneously as well, hence only \eqref{eq:SPCSCAN} with the search of the two least reliable LLRs is taken into account. 
As in \cite{FFastSC}, we assume that the minimum search operations need one clock cycle; thus, the decoding latency of an SPC node requires $L=2$ clock cycles. 

The latency computation for SSCAN in case of REP nodes is symmetrical to that of SPC nodes, leading to the same latency $L=4*(t-1)+2$.
For fast-SCAN, the sum of LLRs in \eqref{eq:rep} takes 1 cycle \cite{FFastSC}, while removing one LLR value takes an additional clock cycle, resulting in $L=2$ clock cycles. 

Type-I nodes have only 2 information bits in the last channels; a node of size $2^t$ is mostly composed of zero-latency rate-0 nodes. 
Hence, type-I is very similar to the REP node also in terms of latency, the only difference being the edge connecting the size-2 rate-1 node to its parent node. Consequently, SSCAN requires $L=4*(t-1)-4+2=4*(t-2)+2$ clock cycles. In fast-SCAN, the even-indexed and odd-indexed bits are interpreted as two independent REP nodes that can be decoded in parallel, reducing the latency to $L=2$.

Similarly, a type-III node of size $2^t$ can be seen as the juxtaposition of two SPC nodes, one involving the even-indexed LLRs, the other involving the odd-indexed LLRs. 
For SSCAN, the latency is reduced to $L=4*(t-2)+2$ as for Type-I, while fast-SCAN decodes the two SPC nodes in parallel, reducing the latency to $L=2$ clock cycles.

\section{Results}
\label{sec:num}


In this Section, we consider polar codes defined in the 5G standard \cite{5G}; we analyze the frequency of occurrence of the identified special nodes, and provide the consequent speedup with respect to standard SCAN decoding. 

\begin{figure}[t!]
  \centering
  \resizebox{0.485\textwidth}{!}{\includegraphics{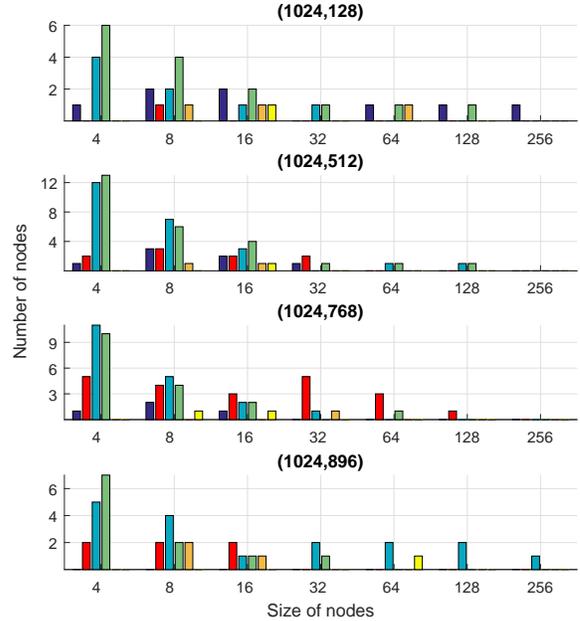}}
  \caption{Number and size of each node type for $N=1024$ and $K=128, 512, 768, 896$. Dark blue, red, light blue, green, orange and yellow are respectively corresponding to rate-0, rate-1, SPC, REP, type-I and type-III nodes.}
  \label{fig:size_number_nodes}
\end{figure}

Figure \ref{fig:size_number_nodes} shows the number and the size of the identified nodes for a given polar code of length $N=1024$ and dimensions $K=128, 512, 768, 896$. 
The nodes are counted considering the maximum size possible. 
For instance, an SPC node of length 128 followed by a rate-1 node of length 128 is counted as an SPC node of length 256.
Low-rate polar codes are more likely to have long rate-0, REP, and Type-I nodes. 
For example, the (1024,128) polar code has a REP node of length 128 and a rate-0 node of length 256. 
Around rate 1/2 the nodes are more evenly distributed, with many nodes having size $\leq16$. 
In high-rate polar codes, the longer nodes are rate-1, SPC and Type-III nodes.  
According to the four rates depicted in Figure \ref{fig:size_number_nodes}, type-I and type-III are the least likely to occur.

\begin{table}
  \caption{Latency and Gain of Fast-SCAN against SCAN for various code lengths and rates}
	\centering
	\tiny
	\begin{tabular}{c|c|c|c||c|c|c|c}
		Polar & Latency & Latency & Gain & Polar & Latency & Latecny & Gain \\
		code & SCAN & fast-SCAN &  (\%) & code &SCAN & fast-SCAN & (\%)\\
		\hline \hline
		 (128,16) & 762 & 50   & {\bf 93.4} & (256,32)   & 1530 & 142 & {\bf 90.7}\\
		(128,64)  & 762 & 146  & {\bf 80.8} & (256,128)  & 1530 & 258 & {\bf 83.1 }\\
		(128,96)  & 762 & 142  & {\bf 81.4} & (256,192)  & 1530 & 194 & {\bf 87.3 }\\
		(128,112) & 762 & 50   & {\bf 93.4} & (256,224)  & 1530 & 186 & {\bf 87.9} \\
		\hline \hline 
		 (512,64) & 3066 & 270 & {\bf 91.2} & (1024,128) & 6138 & 406 & {\bf 93.4}\\
		(512,256) & 3066 & 442 & {\bf 85.6} & (1024,512) & 6138 & 738 & {\bf 88.0 }\\
		(512,384) & 3066 & 354 & {\bf 88.5} & (1024,768) & 6138 & 694 & {\bf 88.7 }\\
		(512,448) & 3066 & 302 & {\bf 90.2} & (1024,896) & 6138 & 338 & {\bf 94.5}\\
	\end{tabular}
\label{tab:latency}
\end{table}

The decoding latency of a sample of 5G-NR polar codes under SCAN and fast-SCAN is presented in Table~\ref{tab:latency}. 
It can be seen that fast-SCAN can reduce the latency of SCAN of more than $80\%$ at code rate $1/2$, regardless of the code length; moreover, at high and low code rates, where special nodes of larger sizes are present, the gain can surpass $94\%$.

\begin{figure}[t!]
  \centering
  \resizebox{0.485\textwidth}{!}{\includegraphics{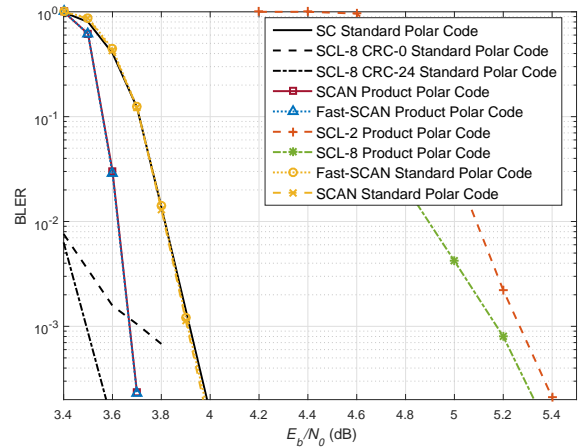}}
  \caption{BLER of product and standard polar code with $N=256^2$ and $K=239^2$.}
  \label{fig:ppcsimu}
\end{figure}

In Figure \ref{fig:ppcsimu}, we consider an additive white Gaussian noise  channel with binary phase-shift keying modulation and provide simulation results for an $N=256^2$, $K=239^2$ polar code under SC, SCL (with and without CRC), SCAN and fast-SCAN. We also provide simulation results of the product polar code scheme presented in \cite{PPC}, where the component codes of the product codes are polar codes. 
Being an iterative concatenated scheme, this scenario can benefit from the soft-in soft-out capabilities of SCAN-based decoding algorithms.
Simulations decode the component polar codes with SCL, fast-SCAN and SCAN decoders; these decoders exchange soft information between iterations, following the scheme detailed in \cite{PPC} in case of SCL. 
SCAN and fast-SCAN consider one internal iteration.
The component code is the $N=256$, $K=239$ code shown on Figure~\ref{fig:Fast256239}. 
We can see that SCAN and fast-SCAN yield the same BLER for both standard and product polar codes, showing that the proposed fast decoding techniques are exact and do not incur any performance degradation. 
SCAN and fast-SCAN slightly improve on the performance of SC for standard polar codes: this is because the hard decision is taken on the a-posteriori information, i.e. the combination of the soft output and the channel LLRs, as would a turbo decoder. 
Fast-SCAN outperforms SCL decoding of product polar codes, yielding a gain of almost 1.5dB at BLER=$10^{-3}$. 
It also outperforms standard polar decoding using SC and non-CRC aided SCL, while it approaches the BLER of CRC-aided SCL. 

For a single internal iteration, SCAN decodes the component code in $6\cdot(2^{8}-1)=1530$ clock cycles, while fast-SCAN needs $58$ clock cycles. The latency of the product polar codes multiplies the latency of the component decoders by the number of row/column half-iterations, since row and column decoding cannot be run in parallel to enable the exchange of information \cite{PPC}. Hence, the maximum latency results in $8*(58+58)=928$ clock cycles with fast-SCAN and $(1530+1530)*8=24480$ clock cycles for SCAN. The standard polar code is decoded in $2N-2=131070$ clock cycles with SC and $2N-2+K=188191$ with SCL, while fast-SCAN requires $12190$ clock cycles.
\section{Conclusions}
\label{sec:conclusions}
In this paper, we have proposed fast-SCAN, a reduced latency decoder based on the SCAN decoding algorithm. 
Fast-SCAN decodes constituent nodes exactly without the need to explore the decoding tree, and thus has no impact on the error-correction performance of SCAN.
At the same time, it can reduce the decoding latency of SCAN of up to $94.5\%$.


\bibliographystyle{IEEEbib}
\bibliography{references}

\end{document}